\documentclass{vgtc}                
\ifpdf
  \pdfoutput=1\relax                   
  \usepackage{graphicx}                
  \DeclareGraphicsExtensions{.pdf,.png,.jpg,.jpeg} 
\else
  \usepackage{graphicx}                
  \DeclareGraphicsExtensions{.eps}     
\fi%

\graphicspath{{figures/}{pictures/}{images/}{./}} 

\usepackage{microtype}                 
\PassOptionsToPackage{warn}{textcomp}  
\usepackage{textcomp}                  
\usepackage{mathptmx}                  
\usepackage{times}                     
\usepackage{cite}                      
\usepackage{tabu}                      
\usepackage{booktabs}                  

\onlineid{0}

\vgtccategory{Research}

\title{Visualization Improvisation}

\author{Swaroop Panda, Shatarupa Thakurta Roy\\
\scriptsize Indian Institute of Technology Kanpur}

\abstract{Teaching visualization design involve making students familiar and make them work with visualization models, framework and perspectives. Visualization research accommodates a plethora of perspectives emerging from researchers of varied backgrounds. These diverse range of perspectives  give rise to multiples models, frameworks and perspectives to teach visualization design. In this paper, we look at an approach to visualization teaching by using improvisational techniques. The basic idea is to design a visualization without using an existing predefined model. Since improvisation, by definition, is not a model or a framework, this work presents a reflection on how improvisation can be a way of teaching visualization design.       
} 


\CCScatlist{ 
 \CCScat{Human-centered computing}%
{Visualization theory, concepts and paradigms}{};
}

\teaser{
  \centering
  \includegraphics[width=\linewidth, height = 5cm]{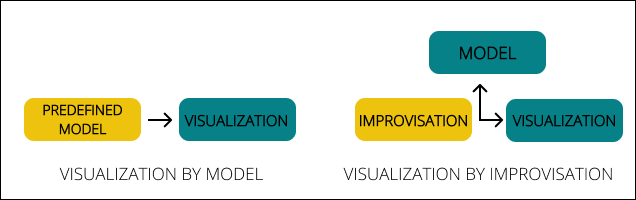}
  \caption{Visualization design by improvisation as compared to visualization design using a predefined model.}
  \label{fig:teaser}
}

\vgtcinsertpkg

\begin{document}



\maketitle

\authorfootertext{\textbf{
This paper has been peer-reviewed and accepted to VisActivities: IEEE VIS Workshop on Data Vis Activities to Facilitate Learning, Reflecting, Discussing, and Designing, held in conjunction with IEEE VIS 2020, Salt Lake City, UT. Workshop organizers: Samuel Huron, Benjamin Bach, Uta Hinrichs, Jonathan C. Roberts, Mandy Keck, http://visactivities.github.io}
}

\section{Introduction}
Visualization has become a very extensive field of research. A lot of researchers with a variety of research interests engage with visualization; they submit and participate in visualization conferences, teach visualization design, practice visualization in industry or work with dedicated visualization research groups within companies. This wide-ranging engagement with visualization has resulted in emergence of a vast literature on visualization, which is evident from not only the number of papers published at visualization, HCI, graphics conferences but also the large number of books, blogs and popular articles on data visualization that have surfaced on the internet. This also signifies the emerging value and importance of visualization in research, communication, presentation and other aspects of digital media.

Given such a setting, how does one go about teaching visualisation design? There are a diverse collection of papers, books, blogs that consider visualization in completely different contexts. As an instance, take the fact that visualization researchers include people from backgrounds such as graphic design, computer graphics, human-computer interaction, data scientists and architecture. Thus, such a varied background of researchers lead to diverse visualization perspectives. A visualization design course may take an algorithmic approach (prioritizing algorithms and software for visualization design), or be more focused on the visual layout (visualization as graphic design) or maybe data-centered (visualization is just a technically correct representation of data). None of these are more or less accurate than the others; but are the consequence of the diversity of visualization design research. 

Reiterating, visualization as a subject, is approachable by a plethora of perspectives. This also has implications on how visualization is deployed in a given setting. For example, take the instance of technique-driven visualization and problem-driven visualization. Technique driven visualization are algorithms that accomplish the task of visualization development; such as graph-drawing or dimensionality reduction. Problem-solving approaches include solving domain-based problems using visualization. These include the use and deployment of visualizations in genomics, business intelligence and journalism among others. So these approaches are means to an end; the end being designing and developing visualizations. In this mishmash of perspectives, how does one introduce the idea of visualization? By a visualization technique? By a visualization to solve a domain problem? Or a visualization for a task? 

In this paper we present improVISe; visualization design by improvisation.It is basically based on the ideas of improvisation. Improvisation, basically, is the idea of creating spontaneously, on the spur of the moment without any advance planning. ImproVISe can serve as a beginner's guide for visualization design, as we shall see, can also act as guideline for experts to explore new models of visualization. ImproVISe, by definition, is not a model or a guideline, but rather a way of doing a given thing (designing visualization in this case). This paper, thus, does not present a framework but rather a reflection of how fundamental ideas of improvisation can be useful in visualization. A reflection, we suppose, is a better way to deal with an open-ended construct such as improvisation.


\section{On Improvisation}
Improvisation, basically is the activity of doing or making a particular thing without any plan by using the available resources. The work by Gary Peters \cite{peters2009philosophy} presents a philosophy of improvisation. Peters get around loosely defining improvisation as \textit{"a work produced within a restricted time frame, within a delimited productive space with delimited resources"}. Improvisation originated from the performing arts where it was the act of performing without any script in hand. For example, musical improvisation involves composition on the spot. Improvisation is the defining attribute of the popular jazz music genre. These skills and features of improvisation have been eventually transported to other fields like the sciences, engineering and other academic disciplines. One of such disciplines, where improvisation has been transported includes design. The work by Elizabeth Gerber \cite{gerber2007improvisation} suggests improvisational principles and techniques for design and observes that improvisational activities by designers can foster innovation, creative collaboration and learning through errors. The work in \cite{de2013leaving} by De Valk et.al. presents a design approach for open-ended play where they offer interaction techniques for personalized games for children; allowing them to make their own rules and goals. The work in \cite{sirkin2014using} by David Sekin et.al. uses an embedded design improvisation as design research tool by using improvisational ideas in conjunction with embodied design and video prototyping. Similarly, in this work, we try to implement the basic ideas of improvisation in the task of visualization design. 

\section{On Improvising Visualization}

ImproVISe, then, is the act of improvising a visualization. The advance planning, in our case, directly translates to having a predefined model developed in advance. So improVISe as a guide is the practice of visualization design without a predefined model. It is important to emphasize that improvising does not amount to designing a visualization without a model; but rather this is the idea of designing a visualization without a \textit{predefined} model. This implies that improvising a visualization simultaneously develops the visualization and a new model.  

What exactly are these predefined models? There has been extensive work on models in the visualization research. These models are frameworks, theories, conceptual structures that are basically used to design, develop, validate and deploy visualizations. The work by Munzner \cite{munzner2009nested} provides a nested model for visualization design and validation. Similarly the work by Liu et.al. \cite{liu2008distributed} provides a distributed cognition theoretical framework for understanding and appreciating visualizations. Kindlmann et.al. \cite{kindlmann2014algebraic} provides an algebraic process for visualization design while the work by Lau et.al. \cite{lau2007towards}provides an information aesthetics context for visualization. Demiral et.al. \cite{demiralp2014visual} provide a visual embedding as a model for designing and evaluating visualizations. The variety of these predefined models (from using aesthetics to cognitive frameworks to algebraic processes) is evidence for the diversity present in visualization design.   

How feasible is it to design a visualization without using such a predefined model? We suppose, there are two basic approaches to capture all of the large variety of models in visualization research literature; one being a unified model encapsulating all the properties and features and the other our proposed approach of improvising a visualization; letting go of all predefined models. We hope other authors in the community will be motivated to pursue the former approach. Also, we suppose letting go of all models also allows to explore more possibilities and push disciplinary boundaries by realizing new models of visualization design that had not been yet explored.  

There is always an underlying model (or process) in any given visualization design process. Thus, an act of improvising a visualization simultaneously designs a visualization and develops a model for visualization design. Also, by improvisation, each model is conditioned by the corresponding visualization. This apparently looks contrary to existing predefined models (upon whom the visualizations are conditioned and thus the models can be used to design visualizations in different contexts). We suppose that models produced by the act of improvisation can also be useful as a predefined model for designing visualization in other unrelated context. However, such a proposition needs to be validated by rigorous empirical research which we hope is an interesting future work for visualization researchers.  

The idea of improvisation is also helpful given the fact that there are a large number of tools and software that are available for visualization design. Data Scientists and software developers use variety of softwares from Tableau, MS Excel to Flourish and other chart generating engines to build visualizations. In fact software like Tableau and MS Excel are coded models for visualization design (where the input is data and the output is a visualization). In teaching and learning visualization design, how do students decide what software to choose for a particular visualization problem. This is an acute issue more so because there are researchers and developers who are dedicated research interests to building new software for visualization design. Thus, with the increasing software and varied tools, improvisation as a way of designing could provide for answer as to which software works the best for the visualization. This is because lack of predefined models allows liberty to the designer to choose software (or a collection of multiple software) in accordance to the desired visualization. The software is not imposed by a predefined model but rather chosen so as to adapt (or conditioned) to the visualization. 

Improvisation does not stop after the visualization is designed. Once the improvisation is done, it is required to place the visualization in context in the existing visualization literature. This is the vital aspect of visualization improvisation; placing the design in context. Take for instance, an existing visualization was improvised by using new data and doing some changes in the color patterns and say this visualization turned out to be highly effective. The important questions that are needed to be asked include why the certain colour palette worked instead of the other? Is there some logic behind it or was it a matter of chance and context? Why did the new data fit the existing visualization? Does all such data-structures fall into the existing visualization? The answers to these questions help comprehending the design process and thereby also helping in the construction of the model. Placing the model in context of other predefined models in existing literature results in gaining more information about its usefulness which we suppose, can help in deploying the model for designing visualization in other different contexts.   


Another crucial aspect of visualization improvisation remains the improviser, that is, the visualization student. What did the student learn? The improvisation method did not teach any existing models or frameworks, nor did it teach how to solve problems using visualizations. Improvisation, as a way of doing, helps the student accomplish the task of visualization design. Improvisation does not feed models, theories, frameworks directly to the students, but rather keeps these peripheral relative to the central aspect to teaching students to design visualization. This is more of an indirect way of teaching and learning visualization design where the student learns existing methods by contextualizing, examining, placing in context and reflecting the process in hindsight. How effective would be improvisation as a teaching tool as compared to other direct methods would be indeed a require larger long term studies. This is not to say that visualization by improvisation can only be useful for beginner students. Improvisation, which is letting go of all predefined models, can also prove effective for experts focusing on exploring further boundaries of visualization design.  


\section{Implementation}
Visualization improvisation is best taught to students in a classroom setting. The curriculum is more project than instruction-based with reading sessions of visualization literature research(as literature reading is important to place the design in context). The instruction is largely technical; based upon teaching fundamentals of software and other technical tools available for visualization, lessons of data, statistics and lessons based on the concerned domain (genetics or aerospace or finance). This is followed by sample demonstrations of improvisation by the instructor. A demonstration could as simple fitting in new data on an existing visualization and modifying colours using a given tool. These demonstrations could nudge the students to further modify existing visualizations with old/new data, change shapes or colours and trying whatever works to make the visualization more effective and impactful. In these instructions, it remains important for the teacher to keep the gist of the improvisational visualization intact;learning to develop and design the desired visualization being the primary and studying models and other existing literature secondary and indirect.


\begin{figure}[h!]
 \centerline{
 \includegraphics[width= 7cm]{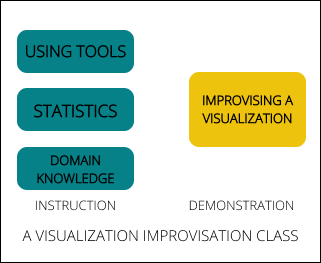}}
 \caption{Components of a visualization improvisation class}
 \label{fig:tm81}
\end{figure}


Visualization improvisation can be also taught by books. However, this is a much more difficult task to accomplish; though it has been done in other fields. Johnstone's book \cite{johnstone2012impro} provides an account of improvisation in theatre, though in retrospect. These books should ideally be able to convince readers to design visualizations by improvising, by following similar methods of technical instructions and improvisation demonstrations. But such books, we suppose, will possibly be much less effective than a classroom engagement with students on this topic.






\section{Conclusion}
Teaching visualization, we have realized, depends a lot on the academic background and style of the concerned instructor. A data scientist would teach visualization much differently than a graphic designer or an HCI researcher. This emerges from the fact that visualization research accommodates multitudes of perspectives that are difficult to disentangle and teach to students. Thus, we propose an improvisational approach to teach visualization design. Improvisation includes visualization design without using any predefined models. We reflected on how improvisation can be an effective and reasonable approach to visualization design; how improvisation process produces new models along with the visualization and we concluded with some suggestions for incorporating visualization improvisation into classroom teaching and with books.    

\bibliographystyle{abbrv-doi}

\bibliography{template}
\end{document}